\documentclass[prb]{revtex4-1}

\usepackage{amsmath}  
\usepackage{amsfonts} 
\usepackage{graphicx} 
\usepackage{caption}
\usepackage{subcaption}
\begin{document}
\title{A DIY Ultrasonic Signal Generator for Sound Experiments}
\date{\today}

\author{Ihab F. Riad}
\affiliation{Physics Department, University of Khartoum, Khartoum, Sudan}
\email{ifriad@gmail.com}
\maketitle
Many physics departments around the world have electronic and mechanical workshops attached to them. The job of these workshops is to design and build experimental setups and instruments for research and the training of undergraduate students.

The workshops are usually run by experienced technicians and equipped with expensive lathing, CNC\footnote{CNC$\equiv$Computer Numerical Control router} machines, electric measuring instruments, and several other essential manufacturing tools.

However, in developing countries such as Sudan the lack of qualified technicians and adequately equipped workshops hampered efforts by these departments to supplement their laboratories with home built equipment. The only other option is to buy needed equipment and experimental setups from specialized manufactures. The latter option is not feasible the departments in developing countries where funding for education and research is scarce and very limited as equipments from these manufactures are typically too expensive.

While our physics department is the best established in Sudan never the less we struggle significantly in equipping our undergraduate teaching laboratories with the needed equipments.

In the past couple of years and with the advancement in prototyping tools like Arduino, and microcontroller development boards, more friendly programming languages, 3D printers, desktop CNC machines and easy to use CAD/CAM softwares the need for highly qualified technicians and expensive workshop equipment was then relaxed \cite{nature-flypi}. The availability of such affordable, and relatively easy to use tools had the effect of increasing the number of teachers and researchers willing to develop their own experimental setups\cite{polis-flypi}.

During the past year and motivated by the need to equip our laboratories with new setups we established a small workshop at the Department of Physics for developing experimental setups to be used in our undergraduate laboratories. The ultrasonic signal generator we are describing here is one such equipment.

\section{Measuring the speed of sound using ultrasonic transducer and receiver}
\begin{figure}
    \centering
    \begin{subfigure}[b]{0.3\textwidth}
        \includegraphics[width=\textwidth]{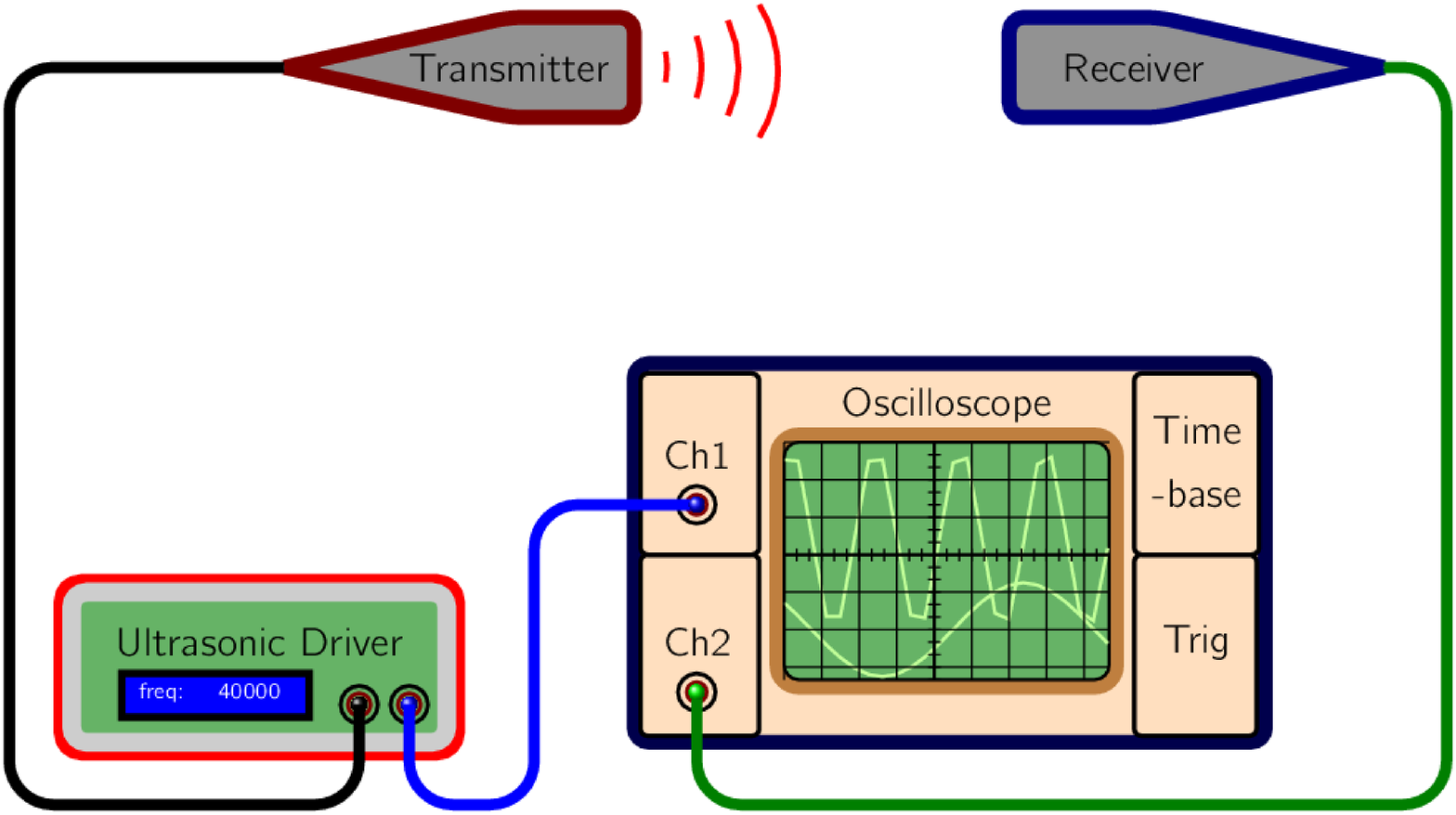}
        \caption{}
        \label{layout}
    \end{subfigure}
    \begin{subfigure}[b]{0.3\textwidth}
        \includegraphics[width=\textwidth]{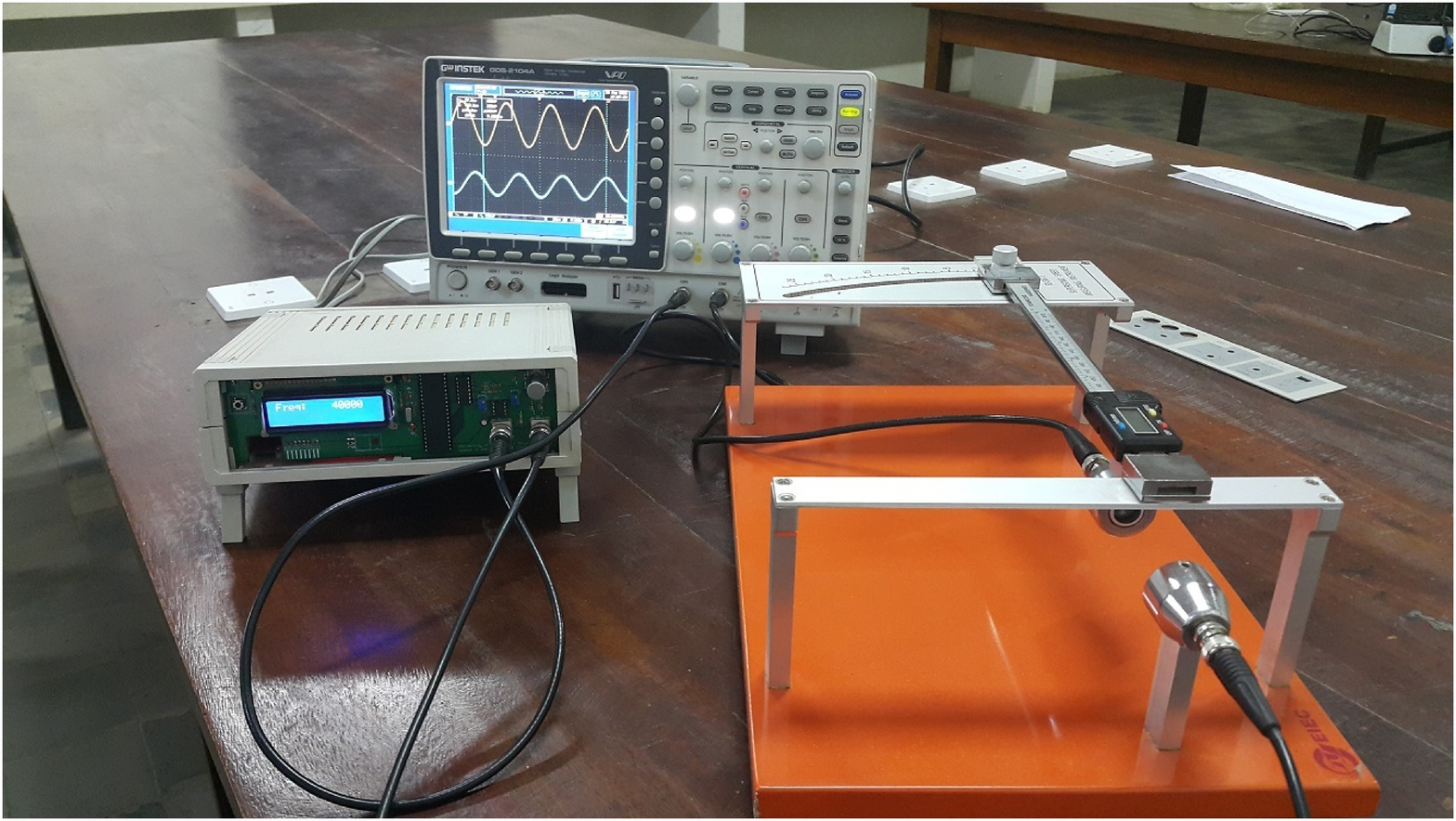}
        \caption{}
        \label{layout1}
    \end{subfigure}
    \begin{subfigure}[b]{0.3\textwidth}
        \includegraphics[width=\textwidth]{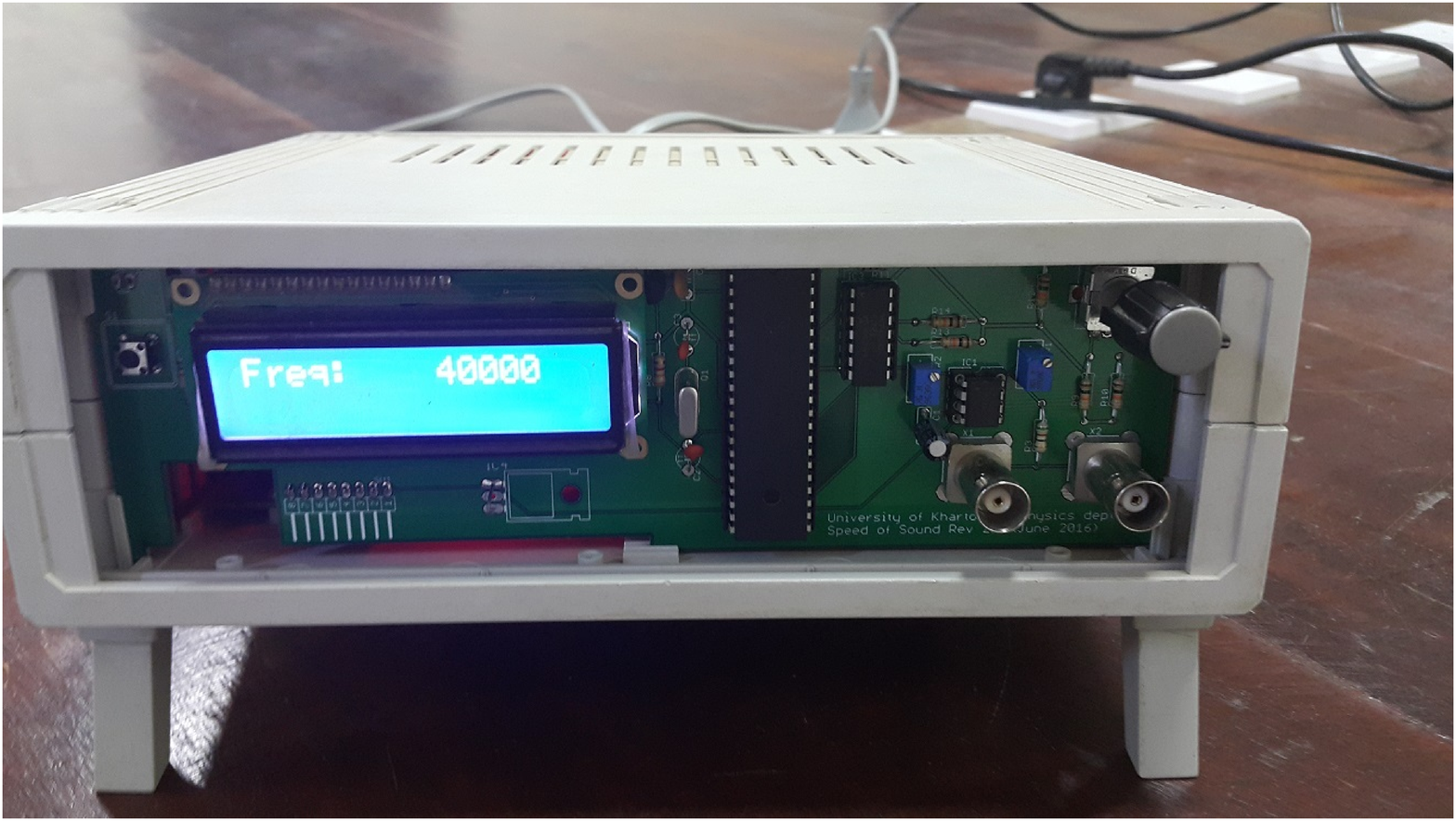}
        \caption{}
        \label{unit}
    \end{subfigure}
    \caption{Panel (a) is the experiment layout, (b) the complete setup, and (c) the ultrasonic function generator.}\label{layouts}
\end{figure}

Measuring the speed of sound using an ultrasonic transducer/receiver and many of its variations is a classic experiment in undergraduate physics curricula \cite{paper1}-\cite{paper9}. It is usually carried out by physics students in their first year of their undergraduate studies. The experiment uses an ultrasonic signal generator and transducer to produce a sine wave in the frequency domain of 35-45khz. The output signal is simultaneously observed in one channel of a two channel oscilloscope. The sound out from the transducer is received by the receiver sitting at a distance $S$ from the transducer (see Figs (\ref{layout} and \ref{layout1}) for the experiment arrangement). The output signal from the receiver is feed into the other channel oscilloscope. Displaying both channels simultaneously on the oscilloscope the time $t$ taken by the sound wave to travel the distance $S$ between the transducer and receiver is measured. The speed of sound is then calculated as
\begin{equation}
 C=\frac{S}{t}.
\end{equation}
Usually measurements are carried out for different values of $S$ and a plot of $S$ Vs. $t$ is used to calculate the speed of sound which in this case is the slope of the plotted straight line.

\begin{figure}
    \centering
    \begin{subfigure}[b]{0.3\textwidth}
        \includegraphics[width=\textwidth]{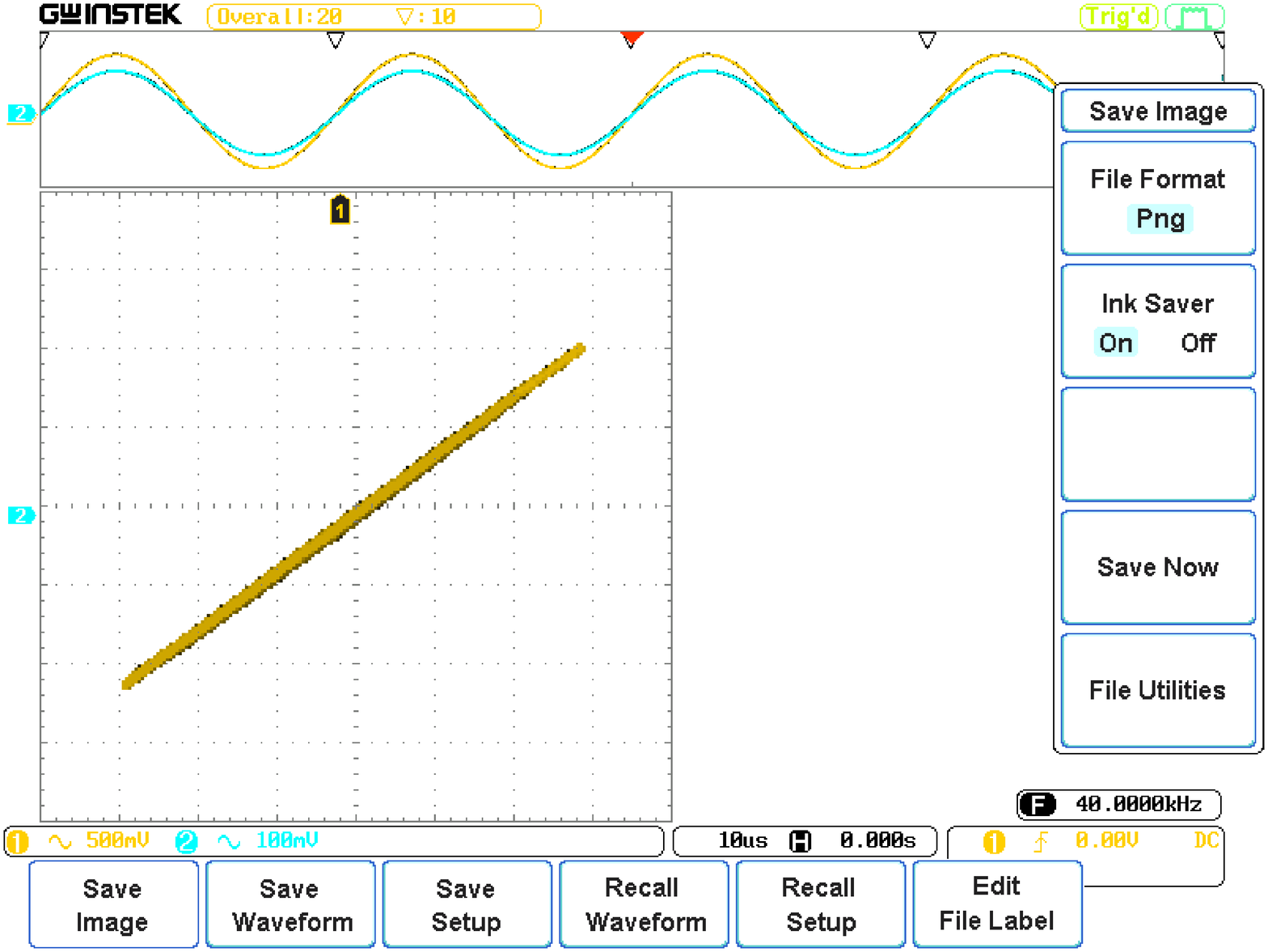}
        \caption{}
        \label{DS0003}
    \end{subfigure}
    \begin{subfigure}[b]{0.3\textwidth}
        \includegraphics[width=\textwidth]{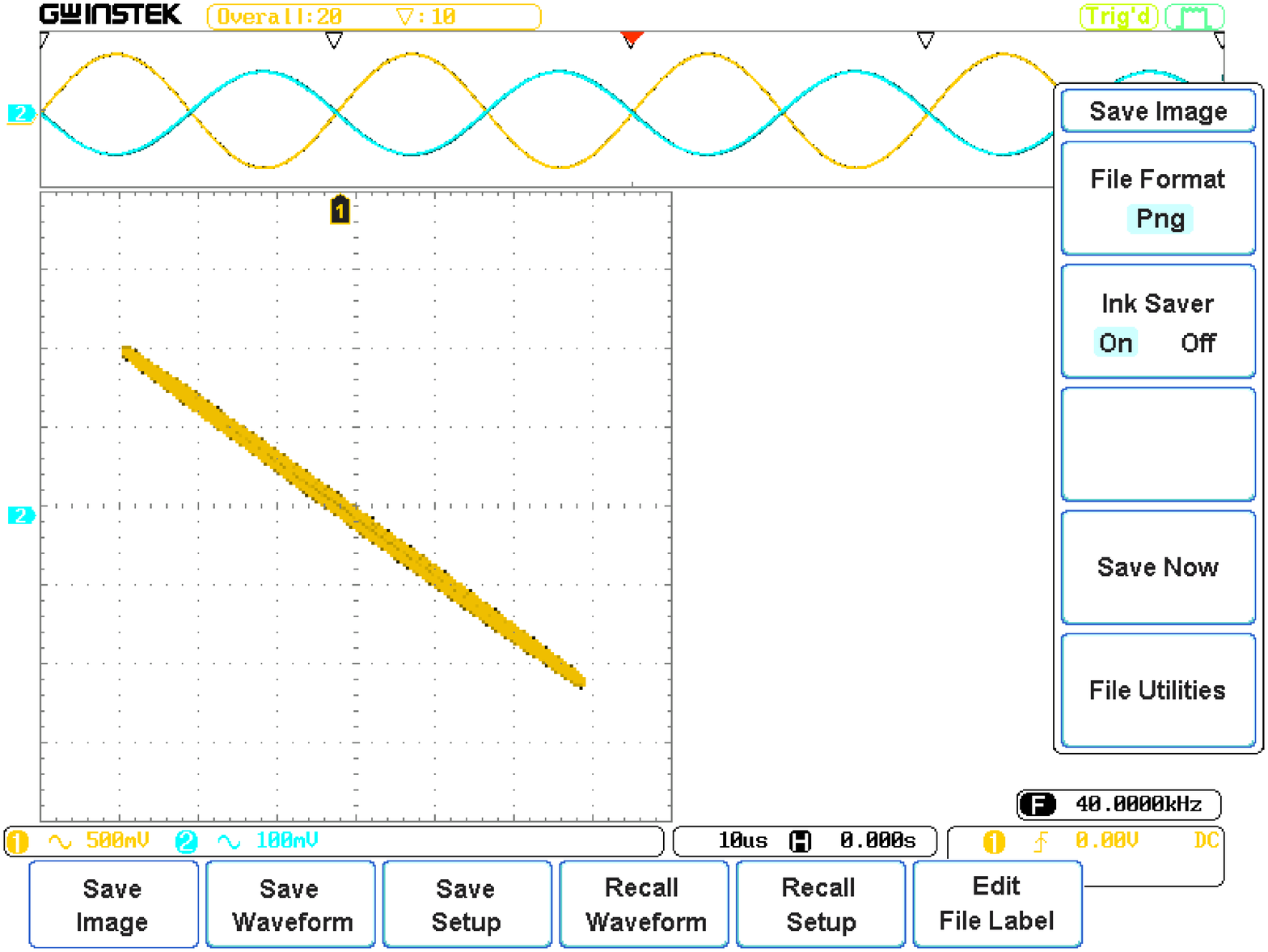}
        \caption{}
        \label{DS0004}
    \end{subfigure}
    \begin{subfigure}[b]{0.3\textwidth}
        \includegraphics[width=\textwidth]{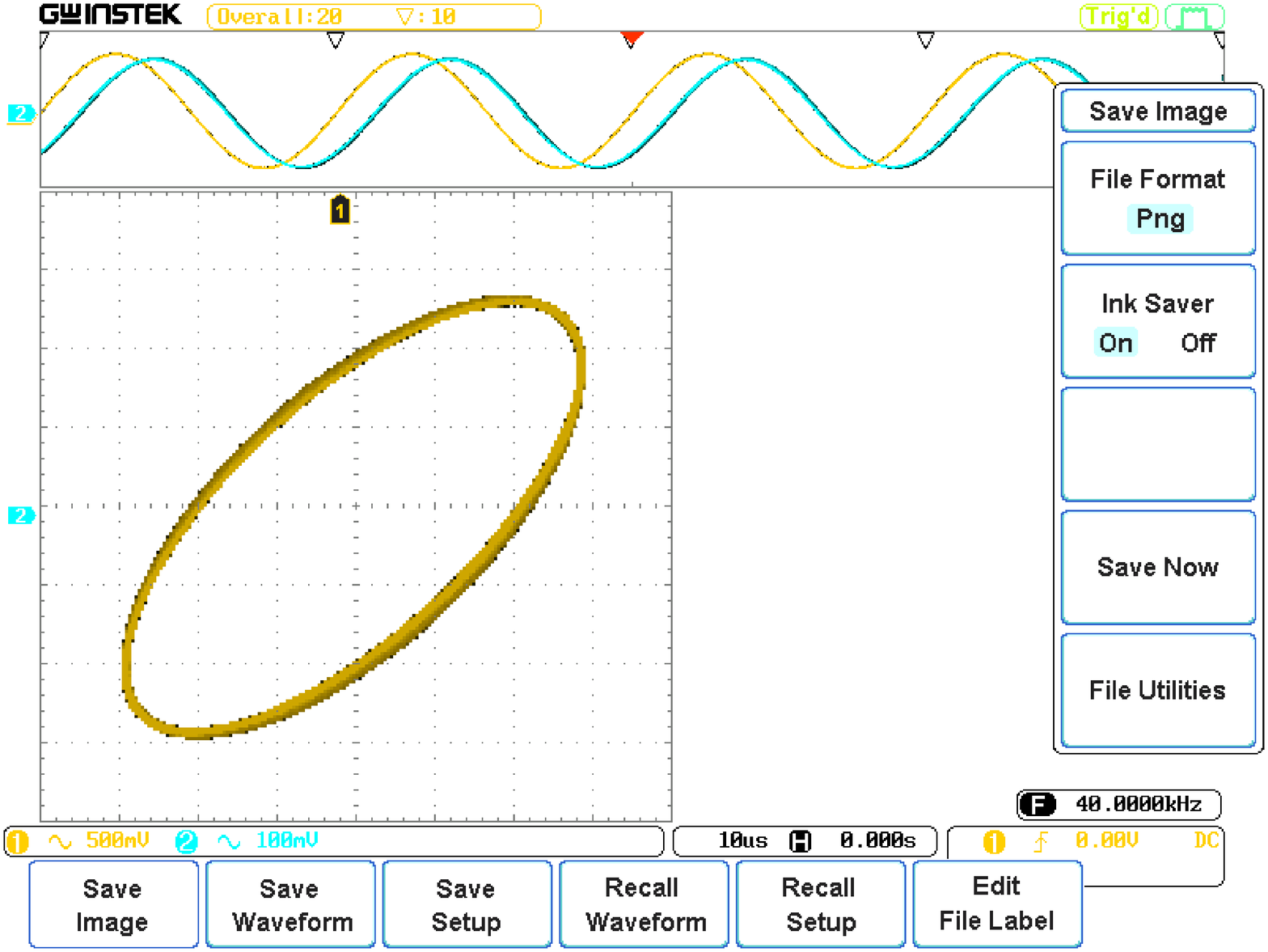}
        \caption{}
        \label{DS0005}
    \end{subfigure}
    \caption{The two channels displayed with the oscilloscope in the xy mode. Panel (a) the tho channels are in phase, (b) the two channels are $180^\circ$ out of phase, and (c) the two channels at a relative phase shift of $\phi$.}\label{fig:animals}
\end{figure}

Another version of the experiment is carried out by setting the oscilloscope into the $XY$ mode. In this mode channel one is the x-axis and channel two is the y-axis. In this setup we observe on form of  Lissajous figures. The shape of the figure gives the phase shift between the signals in the two channels. A straight line with a positive slope indicate zero phase shift between the two signals (\ref{DS0003}), a straight line with a negative slope indicate a $180^\circ$ phase shift (\ref{DS0004}) while an elliptical shape indicate other values for the phase shift (\ref{DS0005}). Moving the transducer and receiver towards/away from each other and noticing the distances giving zero and $180^\circ$ phase shifts the wavelength $\lambda$ of the sound wave could be measured. Knowing the frequency $f$ of the sound wave used the speed of sound is then calculated as
\begin{equation}
 C=\lambda f.
\end{equation}
 \subsection{The Ultrasonic Signal Generator}
\begin{figure*}[!htb]
\centering
\includegraphics[width=0.8\textwidth]{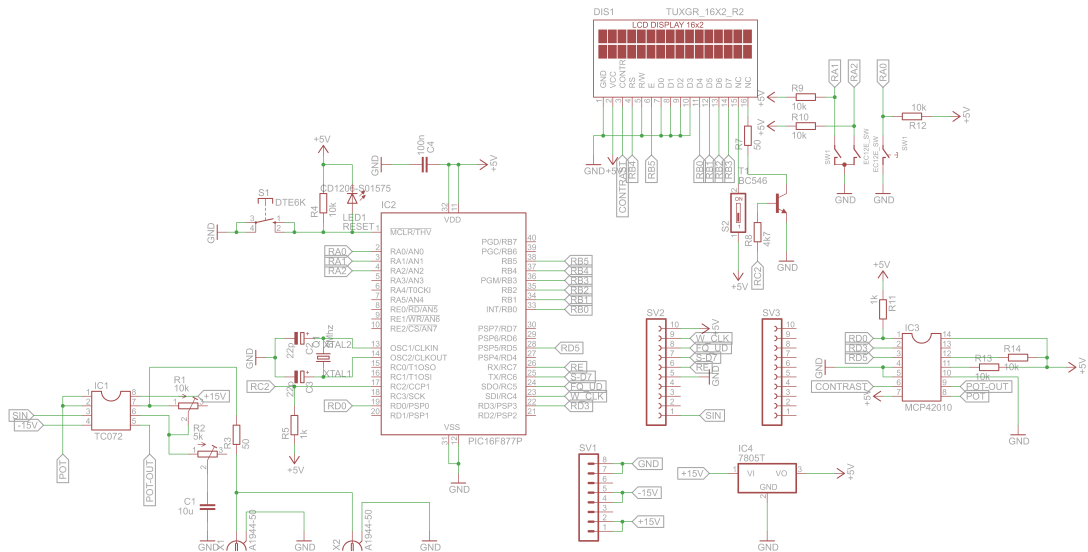}
\caption{Circuit diagram for function generator.}\label{circuit}
\end{figure*}

 In this section we discuss the circuit for the signal generator.  The circuit is build around the AD9850 IC and module \cite{ic}-\cite{module}. The AD9850 is a Direct Digital Synthesis (DDS) IC capable of producing sine  waves in the frequency range of 1hz to  62.5Mhz\footnote{The AD9850 is capable of generation output frequencies of up to one-half the reference clock frequency 62.5Mhz if the reference frequency was 125Mhz}. A module is available \cite{module} that includes the necessary circuitry around the AD9850 to directly connect to a microcontroller, PIC18F4520 in this circuit. A dual digital potentiometer (MCP4210) is used to control the gain of an operational amplifier and the LCD contrast. A rotary encoder with a switch is used to vary the output frequency, amplitude, LCD\footnote{LCD$\equiv$ Liquid Crystal Display} contrast and backlight. The circuit diagram for the signal generator is shown in Fig. (\ref{circuit}). In this version of the generator only a sine wave is output, while in the next iteration we plan to add a square wave and pulse outputs\footnote{The circuit schematics, PCB layout, and gerber files (prepared with Eagle CadSoft) needed to reproduce the PCB are available from the online repository}.

\begin{figure}[!htb]
\centering
\includegraphics[width=0.4\textwidth]{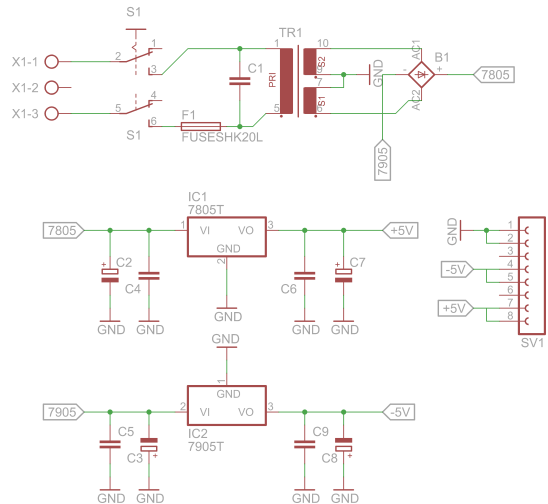}
\caption{Circuit diagram for the unit power supply.}\label{power}
\end{figure}

While this generator is designed to operate in the 40khz frequency domain simple modifications of the microcontroller code\footnote{The code is available online for download. You need the appropriate tools to program the microcontroller} could enable the generator to output frequencies in the range of your choice as long it is within the permissible range of the AD9850.
The power for the main printed circuit board (PCB) comes from another board that output $\pm5$ volts (see Fig. (\ref{power}) \footnote{The schematic and PCB layout for the power board are available online}).  The generator and power boards are both housed in a plastic electronic enclosure.
The generator and power boards could easily be reproduced using commercially available PCB making facilities for example EasyEDA https://easyeda.com/.  Populating the generator with the relevant components only requires average soldering skills and the ability to follow circuit diagrams. It is worth mentioning here that building our function generator gave us a coast reduction of approximately 40\% as compared to the cheapest similar generators available in the market. In addition our generator has the extra benefit that it could be modified to vary the output frequency range.
\section{Conclusion}
The described setup is one of the first units completed in our workshop and it is currently used in our second year laboratory. The above endeavor had demonstrated to us that a good number of experimental setups could be home brewed.
\section{Acknowledgment}
I would like here to I thank the Wellcome Trust Spark Award, in collaboration with TReND in Africa\cite{trend},  for equipping our workshop with a 3D printer and a CNC machine essential in developing the forthcoming equipment.


\begin{thebibliography}{99}
\bibitem{nature-flypi} Elizabeth Gibney (2016) Open-hardware' pioneers push for low-cost lab kit. Nature News (531). doi:10.1038/531147
\bibitem{polis-flypi} Tom Baden, Andre Maia Chagas, Gregory J. Gage, Timothy C. Marzullo, Lucia L. Prieto-Godino, Thomas Euler
 (2015) Open Labware: 3-D Printing Your Own Lab Equipment. PLoS Biol 13(3): e1002086. doi:10.1371/journal.pbio.1002086
\bibitem{paper1} Howard N. Maxwell and Clayton C. Alway, ``A Determination of the Speed of Sound in Air", Am. J. Phys. 18, 192 (1950); doi: 10.1119/1.1932531
\bibitem{paper2} C. K. Manka, ``A Direct Measurement of the Speed of Sound", Am. J. Phys. 37, 223 (1969); doi: 10.1119/1.1975464
\bibitem{paper3} John E. Girard, ``Direct method for measuring the speed of sound", Phys. Teach. 17, 393 (1979); doi: 10.1119/1.2340280
\bibitem{paper4} G. B. Karshner, ``Direct method for measuring the speed of sound", Am. J. Phys. 57, 920 (1989); doi: 10.1119/1.15847
\bibitem{paper5} Rand S. Worland and D. David Wilson, ``The speed of sound in air as a function of temperature", Phys. Teach. 37, 53 (1999); doi: 10.1119/1.880153
\bibitem{paper6} Tony Key, Robert Smidrovskis, and Milton From, ``Measuring the speed of sound in a solid", Phys. Teach. 38, 76 (2000); doi: 10.1119/1.880459
\bibitem{paper7} Mak Se-yuen 2003 ''Wave experiments using low-cost 40 kHz ultrasonic transducers" Phys. Educ. 38 441; doi: 10.1088/0031-9120
\bibitem{paper8} S. Velasco, F. L. Román, A. González, and J. A. White, ``A computer-assisted experiment for the measurement of the temperature dependence of the speed of sound in air", Am. J. Phys. 72, 276 (2004); doi: 10.1119/1.1611479
\bibitem{paper9} Richard E. Berg and Dieter R. Brill, ``Speed of Sound Using Lissajous Figures", Phys. Teach. 43, 36 (2005); doi: 10.1119/1.1845989
\bibitem{ic} \url{<http://www.analog.com/media/en/technical-documentation/data-sheets/AD9850.pdf>}
\bibitem{module} \url{<http://www.eimodule.com/download/EIM377_AD9850_Signal_Generator_Module_V01.pdf>}
\bibitem{trend} \url{<http://trendinafrica.org/>}
\end{thebibliography}
\end{document}